\documentclass[prd,showpacs,preprintnumbers,amsmath,amssymb]{revtex4}
\usepackage{graphicx}
\usepackage{amsmath}
\usepackage{amsfonts}
\usepackage{amssymb}
\usepackage{epsfig,graphics,graphicx}
\begin{document}


%
%
%
%
%



\title{Analytical considerations about the cosmological constant and dark energy}
\author{Everton M. C. Abreu$^a$\footnote{e-mail: evertonabreu@ufrrj.br}, 
Leonardo P. G. De Assis${}^{a,b}$\footnote{e-mail: lpgassis@ufrrj.br} and Carlos M. L. dos Reis${}^{a}$}
\affiliation{${}^{a}$Grupo de F\' isica Te\'orica e Matem\'atica F\' isica, 
Departamento de F\'{\i}sica, Universidade Federal Rural do Rio de Janeiro  \\
BR 465-07, 23851-180, Serop\'edica, Rio de Janeiro, Brazil.\\
${}^{b}$Centro Brasileiro de Pesquisas F\' isicas and Grupo de F\'{\i}sica Te\'{o}rica Jos\'{e} Leite Lopes - GFT JLL \\
P.O. Box 91933, 25685-970, Petr\'{o}polis, Brazil\\
\today}



%
%

%
%
\vskip.2in
\begin{abstract}
The accelerated expansion of the universe has now been confirmed by several independent observations including
those of high redshift type Ia supernovae, and the cosmic microwave background combined with the large
scale structure of the Universe. 
Another way of presenting this kinematic property of the Universe is to postulate the existence of a new and exotic entity, with negative pressure, the dark energy (DE). 
In spite of observationally well established, no single theoretical model provides an entirely compelling framework within which cosmic acceleration or DE can be understood. 
At present all existing observational data are in agreement with the simplest possibility of the cosmological constant to be a candidate for DE.
This case is internally self-consistent and non-contradictory. 
The extreme smallness of the cosmological constant expressed in either Planck, or even atomic units means only that its
origin is not related to strong, electromagnetic and weak interactions. 
Although in this case DE reduces to only a single fundamental constant we still have no derivation from any underlying quantum field theory for its small value. 
From the principles of quantum cosmologies, it is possible to obtain the reason for an inverse-square
law for the cosmological constant with no conflict with observations. 
Despite the fact that this general expression is well known, in this work we introduce families of analytical solutions for the scale factor different from the current literature. 
The knowledge of the scale factor behavior might shed some light on these questions since the entire
evolution of a homogeneous isotropic Universe is contained in the scale factor.
We use different parameters for these solutions and with these parameters we stablish a connection with the equation of state for different dark energy scenarios.
\end{abstract}

\maketitle

\newpage

\section{Introduction}

The research about the cosmological scale factor has attracted intense attention during the last years as well as the investigation about the connection between the cosmological constant and the dark energy issue.  
As the vacuum has a non-trivial role in the early Universe, a $\Lambda$-term in the Einstein field equations is generated.  
This $\Lambda$-term leads to the inflationary phase \cite{weimberg}.  
Based on the inflationary cosmology we can say that during an early exponential phase, the vacuum energy was a large cosmological constant.  
However, the current observed small value of the cosmological constant makes us to assume that $\Lambda$, representing  the energy density of the vacuum, is a variable dynamic degree of freedom which being initially very large went down to its small present value in an expanding Universe \cite{av}.
The last one can be measured through the discrepancies between the infinitesimal value that the cosmological constant has for the present universe (it is very small in Planck units) and the values expected by the Standard Model \cite{berman}.

There is an extensive literature that show examples of phenomenological $\Lambda$-decay laws.  To mention some ideas, the studies comprise a $\Lambda$ depending on temperature, time, Hubble parameter and scale factor \cite{oc}.  A dynamically decaying cosmological constant with cosmic expansion has been considered by several authors \cite{cw,ot,fafm,gasperini,clw,berman,kwe,abdel,beesham}.

The importance of the subject resides in the fact that any nonzero value of $\Lambda$ introduces a length scale and a time scale into the theory of general relativity \cite{bousso}.  The cosmological constant perturbs spacetime dynamics, although the general relativity works on scales much larger than the Planck scales.  However, vacuum quantum fluctuations of the Standard Model rekindle the theory of the cosmological constant problem.   Moreover, the cosmological constant is one of the candidates to be a DE model, namely, with the equation of state $\omega=-1$.  A plaudible DE model is very important to explain the observations introduced by Supernovae (SNe) Type Ia \cite{lista}.  Other candidates are scalar fields models such as quintessence, K-essence, tachion, phantom and etc..  For a review see \cite{lista4}.

Albeit there are many references about a scale factor inverse-square law for the cosmological constant there is a lack of information about the explicit form of the function of the scale factor that obeys this requirement.  In reference \cite{abdel} there are some analytical solutions. 
However, those were obtained for a perfect-fluid matter energy-momentum tensor.  Besides, it is our opinion that there is not an appropriate analysis in the literature that includes the scale factor analytical solutions in a DE approach.  The underlying motivation to carry out these solutions is the fact that the time variation of $\Lambda$ can lead to a creation of matter (i.e., $\Lambda$(t)CDM models) and/or radiation such as to help us in the investigation of the age of the Universe. 

In this work we will use the requirement of the energy-momentum tensor conservation in order to introduce some new analytical solutions for the scale factor different from the published and unpublished \cite{pk} literature.
While the current observations clearly pinpoint to a flat Universe, in the name of generalization,  we will find solutions that have different curvatures, $k=0,-1,+1$ as parameters, i.e., for flat, open and closed universes respectively. 
The obvious interest in a flat $(k=0)$ cosmological constant model has its motivation rooted in the fact that a $\Lambda$-term helps to connect inflation with observations.  
Moreover, with $\Lambda$, it is possible to obtain, for flat Universes, a theoretical age in the observed range, even for a high value of the Hubble parameter \cite{sw,cpt}, but it is not the objective of this work.  With these parameters we analyze some DE models in the background of these analytical calculations.  
In this paper we also assume, as will be said below, that the parameter of the DE equation of state, $\omega$, is constant.  This parame4ter measures the ratio between the dark energy pressure and its energyh density and is its precse value is one of the main tarets in general relativistic cosmology.

In the next section we will depict the field equations and the differential form of the scale factor equation.  In section III we will present some motivations about the reasons to study this specific form for the cosmological constant.  We present some possible analytical solutions for the scale equation in section IV.  
In section V we carry out a very brief review about DE and we analyze its connection with the obtained results.  The final discussions are left for the final section.

\section{Dark energy}

The observations showed that the Universe was not only in expansion but that this expansion is not uniform, as believed by the scientific community, but it is accelerated,  One evidence of this acceleration is through the Hubble-Sandage diagram that shows the observed brightness of the SNe type Ia as a function of the redshift.  It suggest that the bulk of the energy density in the Universe is repulsive and appears like an exotic form of energy with negative pressure.  This exotic energy was dubbed as dak energy and it is considered as an additional component to the ordinary dark energy.  There are many good reviews in the literature \cite{lista4}.

The analysis of dark energy phenomena begun with the observations of Supernovae (SNe) type Ia published independently by two different groups \cite{lista} in the late 90's.  As we said before the equation of state paramete $\omega$, is one of the most wanted number in relativistic cosmology today.  One motivation for this quest is that if one could fix $\omega$ to be precisely $-1$, then, arguably, we could identify the DE with the vacuum state of all existing fields in the Universe, namely, the cosmological constant.  On the other hand, if the precise value obtained is $\omega\not=-1$, then we could not rule out the cosmological constant, but we can work on a model of the DE pressure responsible for the current cosmic acceleration as the potential energu density associated with a dynamical scalar field.  In other words, we have the so-called quintessence field if $-1<\omega<1$ \cite{lista9} or a phantom component for $\omega<-1$ \cite{lista10}.  In both cases we have a violation of the strong energy condition ($\rho+3p>0$).  In the case of the phantom component, moreover, we have the violation of the null energy condition ($\rho+p>0$, for a homogeneous and isotropic Universe).  Bounds on the value of $\omega$ can be obtained from observation such as the mentioned magnetic measurements of SNe Ia.
There are other evidences that came with the anisotropies of the Cosmic Microwave Background (CMB) observations \cite{lista2}, the Large-Scale Structure (LSS) data \cite {lista3}, the age of high-z objects \cite{lista5}, measurements of the angular size of radio sources \cite{lista6}, observation of the X-ray luminosity of galaxy clusters \cite{lista7}, gravitational lensing \cite{lista8} and there are others that the interested reader can find in \cite{lista} showing that a standard cold matter scenario is not enough.  Recently, in \cite{da} the authors showed, using the lookback time techique, how to provide bounds on the cosmological parameters with high accuracy.  

In spite of the number of DE models, the cosmological constant and the quintessence model are the most probables candidates.  For the quintessence model, the equation of state is determined dynamically by the scalar field or tachyon.

It is important to notice that the concept of an accelerated Universe and consequently of the existence of a DE component can be considered as an introduction of a new era in general relativity (GR).  It is possibloe that we have to review all the GR concepts since repulsive gravity is possible in GR.  Through this section we will present briefly some possible alternative explanations for the acceleration of the Universe and connect these models with the results obtained in the last section, which is our actual objective.

To accomplish this we have to analyze the main dark energy ingredient which is its equation of state given by equation $p=\omega\,\rho$.  We will see  that the alternative models of dark energy have their origin through different values of $\omega$.  It is worthwhile to stress that it is very plausible to consider this parameter constant because observational data can hardly distinguish between a varying and a constant equation of state \cite{bk}.  But there are many studies that consider this parameter as a function of the redshift $z$ or the scale factor $a(t)$.  Here we will consider this parameter as constant.

From the conservation of the energy momentum tensor through the Bianchi identities we have a continuity equation given by
\begin{equation} \label{22}
\dot{\rho}=-3\frac{\dot{a}}{a}\,(\rho\,+\,p)\,\,,
\end{equation} 
and substituting the equation of state above and solving for $\rho$ we have that
\begin{equation} \label{23}
\rho\,=\,\rho_0\,a^{-3(1+\omega)}\,\,,
\end{equation}
where $\rho_0$ is the density of non-relativistic matter today.  For $\omega=0$ we have that $p=0$ and this scenario is characterized by a dust dominated Universe, i.e., $\rho=\rho_0\,a^{-3}$.  For $\omega={1\over 3}$ we have $\rho=\rho_0\,a^{-4}$, i.e., a radiation dominated Universe.
However, for both cases we have a deceleration expansion of the Universe scenario.  Notice that it is the same result provided by the Newton gravitational theory.

As we claimed in the first sections, the scale factor been proportional to the Universe radius, as said above, describes the growth of the Universe and it is defined so that at the present time $t_0$, $a(t_0)=1$.  It can be proved both classically and relativistically that the time evolution of the expansion obeys
\begin{equation} \label{21}
\frac{\ddot{a}}{a}\,=\,-\frac{4\pi G}{3}\,(\rho\,+\,3p)\,\,,
\end{equation}
where $\ddot{a}(t)$ measures de acceleration of the Universe.

The experiments described above confirm that at the present time, $\ddot{a}(t)>0$, characterizing an accelerating Universe.  This acceleration implies that, either: the Universe is dominated by some particles (dark matter) or by a field (dark energy) that has a negative pressure.  In particular, substituting the equation of state in (\ref{21}) we have the condition $\omega < -{1\over 3}$.  And again we have two situations:  or there is in fact a non-zero cosmological constant; or the GR or the standard model, is incorrect.  For $\omega=-{1\over 3}$ we have an acceleration/deceleration boundary.  The DE is so enigmatic that we can have two different scenarios: if DE is a cosmological constant the Universe will be acceleration indefinitely and will become empty; or if it is a dynamical scalar field, this last may decay and refill the Universe with matter and energy.

Another scenario that appears in the literature is the one of a X-matter plus cold dark matter (the well known XCDM parametrization) where both fluid components are separately conserved.  This cosmological model is characterized by an equation of state $-1\leq\,\omega\,<0$.  The model dominated by the cosmological constant $p=-\rho$, i.e., $\omega=-1$ is a limiting case of the X-matter parametrization.  In terms of energy density we have that, using (\ref{23}), $\rho=\rho_0$, and the energy density is constant in any epoch.  hence, we can say that the cosmological constant corresponds to a fluid with a constant equation of state $\omega=-1$.


\section{The scale factor equation}

Let us begin reviewing some basic concepts.  The Einstein field equations are
\begin{equation} \label{1}
G_{\mu\nu}\,+\,\Lambda\,g_{\mu\nu}\,=\,8\,\pi\,G\,T_{\mu\nu}\,\,,
\end{equation}
where $\Lambda$ has often been treated as a constant of nature and where
$G_{\mu\nu}$ is the Einstein tensor given by
\begin{equation}\label{2}
G_{\mu\nu}\,=\,R_{\mu\nu}\,-\,{\frac{1}{2}}\,R\,g_{\mu\nu}%
\end{equation}
and $T_{\mu\nu}$ is the energy-momentum tensor. We are using relativistic
units (i.e., $c=1$) and we are assuming that the Universe is homogeneous and isotropic.
In this case we will use the Robertson-Walker line element.  Let us consider also a 
perfect-fluid-like ordinary matter with pressure $p$ and energy density $\rho$ \cite{oc}.

Taking the covariant divergence of (\ref{1}) and using Bianchi identities to
guarantee the vanishing of the covariant divergence of the Einstein tensor, it
follows (assuming the energy-momentum conservation law) that the covariant
divergence of $\Lambda\,g_{\mu\nu}$ must vanish also and therefore $\Lambda$
is constant constituting a geometrical interpretation of $\Lambda$.

As usual in recent works, we can move the cosmological term to the right side of (\ref{1}) 
in order to reconstruct the Einstein equation with an alternative form such as,
\begin{equation} \label{3}
G_{\mu\nu}\,=\,8\,\pi\,G\,\tilde{T}_{\mu\nu}%
\end{equation}
where
\begin{equation}
\label{4}\tilde{T}_{\mu\nu}\,\equiv\,T_{\mu\nu}\,-\,\frac{\Lambda}{8\pi
G}g_{\mu\nu} \,\,,
\end{equation}
and we can interpret the cosmological term as part of the matter content of the
universe, rather than a geometrical term. With this effective momentum-tensor
$\tilde{T}_{\mu\nu}$ satisfying the energy conservation as
\begin{equation}
\label{5}\nabla^{\nu}\,\tilde{T}_{\mu\nu}\,=\,0
\end{equation}
there are no reasons in order to refuse a varying cosmological terms
\cite{oc}. At this point we have to observe that since we use $c=1$, we will
not consider here cosmological models with a time varying speed of light
\cite{am}.

The incorporation of the cosmological term in the definition written in (\ref{4}) implies that the effective energy-momentum tensor has the perfect fluid form, with an effective pressure given by $\tilde{p}\equiv p\,-\,\Lambda/8\pi G$ and an energy density as $\tilde{\rho}\equiv \rho\,+\,\Lambda/8\pi G$ \cite{oc}.

Using the field equations (\ref{3}) and (\ref{4}) and the law of energy-momentum conservation, equation (\ref{5}), we can write
\begin{equation}
\label{6}
\dot{a}^2\,=\,\frac{8\pi G}{3}\rho\,a^2\,+\,\frac{\Lambda}{3}\,a^2\,-\,k
\end{equation}
and
\begin{equation} \label{7}
\frac{d}{da}\left[\left(\rho+\frac{\Lambda}{8\pi G}\right)a^3\right]\,=\,-3\left(p\,-\,\frac{\Lambda}{8\pi G}\right)a^2\,\,.
\end{equation}
For the equation of state we can write that 
\begin{equation} \label{8}
p\,=\,(\gamma\,-\,1)\,\rho
\end{equation}
with $\gamma=$ constant.   Note from (\ref{8}) that for $\gamma=1$ we have matter-dominated epoch ($p=0$) and for $\gamma=\frac{4}{3}$ we have radiation-dominated epoch.

Notice that, in order to obtain a convenient form for athe nest equations we constructed the equation of state in (\ref{8}).  In a DE scenario, as said in the last section, we are clearly making $\omega=\gamma-1$.  The reason for this will be clarified in the next lines.

Substituting (\ref{8}) in (\ref{7}) we have that
\begin{equation} \label{9}
\frac{d}{da}(\rho a^{3\gamma})\,=\,-\frac{a^{3\gamma}}{8\pi G}\,\frac{d\Lambda}{da}
\end{equation}
Now, using (\ref{6}) and (\ref{9}) we obtain that 
\begin{equation} \label{10}
\ddot{a}\,=\,\frac{8\pi G}{3}\,\left(1-\frac{3\gamma}{2}\right)\rho a \,+\,\frac{\Lambda}{3}\,a
\end{equation}

Using (\ref{6}) to eliminate $\rho$ in (\ref{10}) (or vice-versa) we can write the equation
\begin{equation} \label{11}
{\ddot{a}\over a}\,=\,\left({\dot{a}}^2\,+\,k \right)\left(1\,-\,{3\gamma\over 2}\right){1\over a^2}\,+\,{\gamma\over 2}\Lambda\,\,,
\end{equation}
which is a differential equation for the scale factor and has a cosmological term.  We clearly see that this equation is independent of the fact that $\Lambda$ is constant or not \cite{oc}.

Our objective from now on is to present analytical solutions for the scale factor in equation (\ref{11}).  But to accomplish this we have to choose a relevant expression for the cosmological term.

As said above, there is a whole literature that approaches this issue (see \cite{oc} and the references therein).  Here we choose a specific one (well known in the literature) but, we produce a variation in the curvature also.

We can summarize the forms of the cosmological constant depending on the scale factor in a general expression as \cite{matyjasek}
\begin{equation} 
\label{11.1}
\Lambda(t)\,=\,\alpha\,a^{-m}\,+\,\beta\left({\dot{a}\over a}\right)^2\,\,,
\end{equation}
where $\alpha$, $\beta$ and $m$ are constants.  Ozer and Taha \cite{ot} and Chen and Wu \cite{cw} concluded, although in different contexts, that the time dependence of $\Lambda$ should have $\beta=0$ and $m=2$ \cite{matyjasek}.  The interested reader can  see the different variations concerning these parameters in \cite{oc}.

Hence, let us choose the cosmological term as 
\begin{equation} \label{12}
\Lambda\,=\,{\cal B}\,a^{-2}\,\,,
\end{equation}
where ${\cal B}$ (we use here the same parameter used in \cite{oc}) is a pure number of order $1$ which should be calculable in a model for the time variation of $\Lambda$.   We can consider that the parameter ${\cal B}$ is a new cosmological parameter to be computed from observations, substituting $\Lambda$.  If ${\cal B}=0$ we have, obviously, an ansatz of $\Lambda=0$.  


\section{Some considerations and motivations}


With a time variation like the exposed in (\ref{11.1}) the values of $\Lambda$ in the early Universe could be much bigger than the present value of $\Lambda$.  However, even that order of small magnitude it was arguably large enough to drive various symmetry breakings that occurred in the early Universe.  With a $\Lambda=0$ ansatz we cannot have such scenario.  Otherwise, if ${\cal B}\not=0$, the equation (\ref{11.1}) can be identified as a ``medium" time variation \cite{cw}.  

For a positive ${\cal B}$ our Universe is flat.  In fact, considerations based on the second law of thermodynamics lead to the restriction ${\cal B}\geq0$ \cite{ot}.  Also, if ${\cal B}\geq k$ the Universe will expand forever; if ${\cal B}<k$ the Universe will collapse in the future.  For $\gamma>1$ in equation (\ref{11}), even a closed Universe will expand forever and the $k=0$ case is no longer critical for the collapse of our Universe \cite{ot}.
It can be shown that if our Universe is flat, observationally ${\cal B}$ should be positive \cite{cw}.  
 
If we write equation (\ref{12}) in a general form, i.e.,  as a function of the scale factor $a^{-n}$, like in the first term in equation (\ref{11.1}), the value $n=2$ has dimensional reasons.  With $n=2$, both $\hbar$ and $G$ disappear and we have, with $c=1$ only, the equation (\ref{12}) \cite{cw}.  The presence of $\hbar$ on the Einstein equations would be rather disturbing.

Nevertheless, equation (\ref{12}) can be obtained theoretically based on general assumptions on quantum gravity.  It can be considered also that this formulation has no conflicts with current data and can furnish some explanation about the inflationary scenario \cite{cw}.  In other words we can say that this model is singular and preserves the standard picture of the early Universe.  Notice that the condition $d\Lambda/da < 0$ requires that ${\cal B}<0$ independently of the curvature index $k$ and hence $\Lambda>0$ for all $t\geq 0$ \cite{av3}. We will not make inflation considerations for the time being.  

In another point of view, the dependence on $a$ used in (\ref{12}) was first introduced by Gasperini \cite{gasperini} in a thermal approach.  An alternative motivation for this kind of dependence is that it also appear in some string dominated cosmologies \cite{string}.   We can also obtain an equation (\ref{12}) form based on semiclassical Lorentzian analysis of quantum tunneling \cite{strominger}.  Although this kind of dependence is only phenomenological and does not come from particle physics first principles \cite{sw} it can be considered as a laboratory in order to investigate the models it generalizes.  As mathematically simple, we can obtain some analytical solutions.

\section{The analytical solutions}
\bigskip

After this brief historical review, let us now substitute (\ref{12}) in
(\ref{11}) to have that
\begin{equation} \label{13}
{\frac{\ddot{a}}{a}}\,=\,\left(  1\,-\,{\frac{3\gamma}{2}}\right)
\,(\dot{a}^{2}\,+\,k)\,{\frac{1}{a^{2}}}\,+\,{\frac{\gamma}{2a^{2}}%
}\,\mathcal{B}\,\,,
\end{equation}
which can be written as in \cite{oc} as
\begin{equation}   \label{14}
a\,\ddot{a}\,+\,\Lambda_{1}\,\dot{a}^{2}\,+\,\Lambda_{2}\,=\,0\,\,,
\end{equation}
where
\begin{align} \label{15}
\Lambda_{1}  &  =-\,\left(  1\,-\,{\frac{3\gamma}{2}}\right)
\,\,,\nonumber\\
\Lambda_{2}  &  =-\,\left(  1\,-\,{\frac{3\gamma}{2}}\right)  \,k\,-\,{\frac
{\gamma}{2}}\,\mathcal{B}\,\,,
\end{align}
and $\gamma$, $k$ and $\mathcal{B}$ constitute a group of free parameters
which have the respective physical considerations mentioned above. In our
analysis, the parameter $k$, as said above, will have the values $0$, $-1$ and
$+1$. Just to remember, if we observe (\ref{12}) we can see clearly that if
$\mathcal{B}=0$, then the expression covers the $\Lambda=0$ case. If
$\mathcal{B}\neq0$ \cite{cw} we have two options, $\mathcal{B}\geq k$ where
the Universe will expand forever or $\mathcal{B}<k$ and then the Universe will
collapse in the future. If $\mathcal{B}>1$, even a closed Universe (with
$k=1$) will expand forever and the $k=0$ case is no longer critical for the
collapse of our Universe \cite{cw}.

For the case $\Lambda_{1}=-1/2$ and substituting in (\ref{14}) we have that
\begin{equation} \label{16}
a\,{\frac{d^{2}a }{dt^{2}}}\,-\,{\frac{1}{2}}\left(  {\frac{da}{dt}%
}\right)  ^{2}\,+\,\Lambda_{2}=0\,\,.
\end{equation}

After some algebraic work we can reduce the solution to a simple form as
\begin{equation} \label{18}
a(t)\,=\,-\,2\,{\frac{\Lambda_{2}}{c_{1}}}\,+\,{\frac{c_{1}}{4}%
}\,(t\,+\,c_{2})^{2}\,\,,
\end{equation}
where $c_{1}$ and $c_{2}$ are constants to be determined. Observe that we have
written the equation above in order to maintain the parameter $\Lambda_{2}$ in
evidence for the following analysis below.

For this case of $\Lambda_{1}=-1/2$, from equations (\ref{15}) we have that
$\gamma=1/3$ and consequently that $\Lambda_{2}$ has a dependence on $k$ and
$\mathcal{B}$ that can be written as
\begin{equation}  \label{18.1}
\Lambda_{2}=-{\frac{1}{2}}\,k\,-\,{\frac{1}{6}}\,\mathcal{B}\,\,,
\end{equation}
and we have that for the values of $k=0,+1,-1$,
\begin{align} \label{18.2}
& \Lambda_{2}^{k=0}  =-{\frac{1}{6}}\,\mathcal{B}\,\,,\nonumber\\
& \Lambda_{2}^{k=-1}   ={\frac{1}{2}}\,-\,{\frac{1}{6}}\,\mathcal{B}%
\,\,,\nonumber\\
& \Lambda_{2}^{k=1}    =-\,{\frac{1}{2}}\,-\,{\frac{1}{6}}\,\mathcal{B} \,\,,
\end{align}
and the analysis about the expansion or collapse of the Universe can be
accomplished as above.

Since the value $\mathcal{B}=0$ can be obviously discarded, we can stress that
the general behavior of the solution (\ref{18}) does not change. As a final
observation we can say for $\Lambda_{1}=-1/2$ the equation of state can be
written as $p=-2/3$, where we have an inflation scenario.

As usual we measured in units of Hubble times ($\tau=H_{0}t$) and we have
imposed the following boundary conditions at present epoch:

\bigskip%
\[
a\left(  \tau_{0}\right)  =1\text{ and }H\left(  \tau_{0}\right)  =H_{0.}%
\]

\bigskip 

After have calculated the expression for the Hubble parameter we used
these boundary conditions to determine the free parameters values in the scale
factor  since we have two equations (scale factor and Hubble parameter) and
two incognites c1 and c2.

To plot the figures of this work we have set $H_{0}$ the $74$ that is inside
of the current accept value of $H_{0}$ ($73$ $\pm10$ $Km$ $s$ $^{-1}$ $Mpc$
$^{1}$) \cite{fre96}.

In figure 1, after the change of variable $\tau =H_{0}t$, we present the plot of the scale factor versus $\tau $ with $\Lambda _{1}=\frac{1}{2}$ and $\Lambda _{2}$ varying from $-20$ to $20$.






\begin{figure}[!h]
\begin{center}
\rotatebox{-90}{\includegraphics[width=0.3\textwidth]{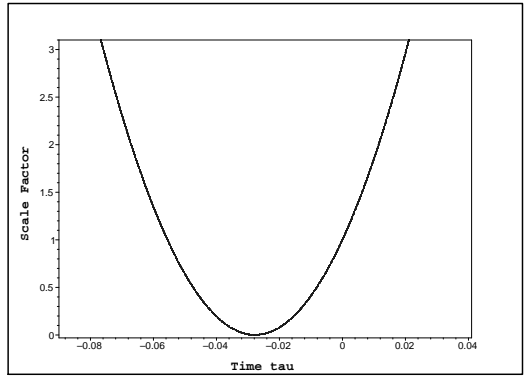}}
\caption{Scale factor, equation (18), versus $\tau $ with $\Lambda _{1}=\frac{1}{2}$ and $\Lambda _{2}$ varying from $-20$ to $20$ (Units $H_{0}^{\text{ \ }-1}$ )}
\end{center}
\label{fig1}
\end{figure}

For $\Lambda_{1}=-1$ we have two general solutions for the equation (\ref{13}),
\begin{align} \label{19}
a_{1}(t)\,=\,{\frac{1}{4c_{1}}} \left[  e^{c_{1}(t+c_{2}%
)}\,-4\,\Lambda_{2}\,e^{-c_{1}(t+c_{2})}\right]  \,\,,\nonumber\\
a_{2}(t)\,=\,{\frac{1}{4c_{1}}} \left[  e^{-c_{1}(t+c_{2})}\,-4\,\Lambda
_{2}\,e^{c_{1}(t+c_{2})}\right]  \,\,,
\end{align}
where the difference is the signs of the exponentials.

For $\Lambda_{1}=-1$ we have a little different behavior. From (\ref{15}) we
have that $\gamma=0$ and consequently that $\Lambda_{2}=-k$, so we can write
that $\Lambda_{2}^{k=0}=0$, $\Lambda_{2}^{k=1}=-1$ and $\Lambda_{2}^{k=-1}=1$.
For $\Lambda_{2}=0$ the solutions (\ref{19}) have different behaviors
altogether, namely,
\begin{align} \label{19.1}
a_{1}(t)  &  ={\frac{1}{{c_{1}}}}e^{c_{1} (t+c_{2})}%
\,\,,\nonumber\\
a_{2}(t)  &  ={\frac{1}{{c_{1}}}}e^{-c_{1} (t+c_{2})}\,\,,
\end{align}
and the equation of state now is $p=-\rho$, again an inflation scenario.

For $\Lambda_{2}=\pm1$ we have that,
\begin{align} \label{19.2}
a_{1}^{k=\pm1}(t)  &  ={\frac{1}{4c_{1}}} \left[  e^{c_{1}
(t+c_{2})}\,\mp\,4\,e^{-c_{1}(t+c_{2})}\right]  \,\,,\nonumber\\
a_{2}^{k=\pm1}(t)  &  ={\frac{1}{4c_{1}}} \left[  e^{-c_{1} (t+c_{2})}%
\,\mp\,4\,e^{c_{1}(t+c_{2})}\right]  \,\,,
\end{align}
and we can see again completely different formulations for the scale factor in
an inflation scenario.

\vspace{1cm}

In figure 2,  we present the plot of the scale factor versus $\tau ,$ first equation (23), with $\Lambda _{1}=-1$ and $\Lambda _{2}=-1$ $\left( k=1\text{ and }\gamma =0\right) .$






\begin{figure}[!h]
\begin{center}
\rotatebox{-90}{\includegraphics[width=0.35\textwidth]{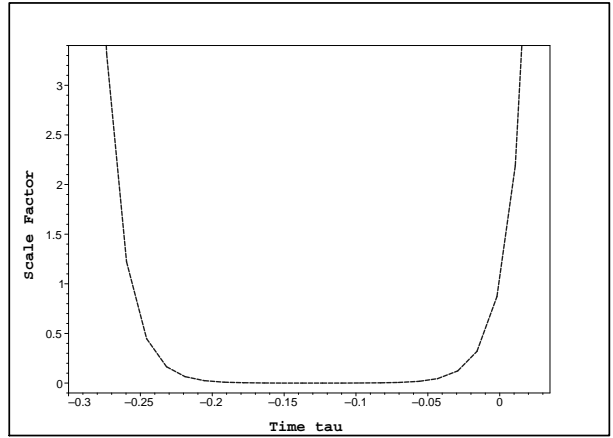}}
\caption{Scale factor versus $\tau ,$ first equation (23), with $\Lambda _{1}=-1$ and $\Lambda _{2}=-1$ $\left( k=1\text{ and }\gamma =0\right)$ (Units $H_{0}^{\text{ \ }-1}$ )}
\end{center}
\label{fig2}
\end{figure}

\pagebreak

In figure 3,  we present the plot of the scale factor versus $\tau ,$ second equation 23, with $\Lambda _{1}=-1$ and $\Lambda _{2}=-1$ $\left( k=1\text{ and }\gamma =0\right).$






\begin{figure}[!h]
\begin{center}
\rotatebox{-90}{\includegraphics[width=0.35\textwidth]{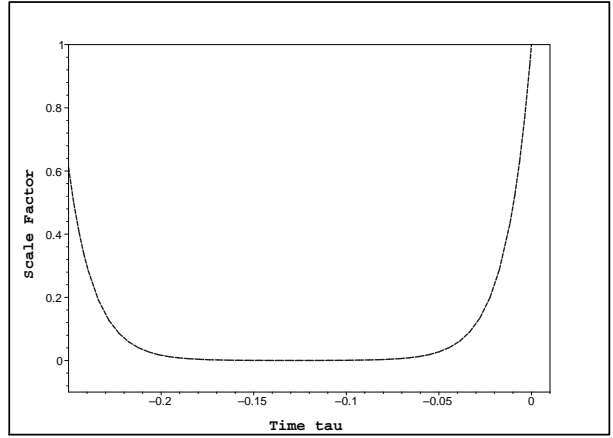}}
\caption{Scale factor versus $\tau ,$ second equation (23), with $\Lambda _{1}=-1$ and $\Lambda _{2}=-1$ $\left( k=1\text{ and }\gamma =0\right)$ (Units $H_{0}^{\text{ \ }-1}$ )}
\end{center}
\label{fig3}
\end{figure}


In the figure 4,  we present the plot of the scale factor versus $\tau ,$ first equation (23), with $\Lambda _{1}=-1$ and $\Lambda _{2}=1$ $\left( k=-1\text{ and }\gamma =0\right)$ with singularity in $\tau $=$-0.1284233275$.






\begin{figure}[!h]
\begin{center}
\rotatebox{-90}{\includegraphics[width=0.35\textwidth]{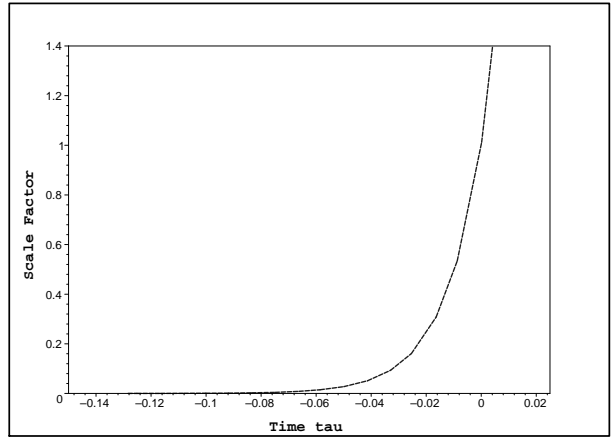}}
\caption{Scale factor versus $\tau ,$ first equation (23), with $\Lambda _{1}=-1$ and $\Lambda _{2}=1$ $\left( k=-1\text{ and }\gamma =0\right)$  with singularity in $\tau $=$-0.1284233275$. (Units $H_{0}^{\text{ \ }-1}$ )}
\end{center}
\label{fig4}
\end{figure}

For $\Lambda_{1}=1$ we have the following solution,
\begin{equation}
\label{20}a(t)\,=\,\pm\sqrt{{\frac{c_{1}}{\Lambda_{2}}}\,-\,\Lambda
_{2}(t\,+\,c_{2})^{2}}\,\,,
\end{equation}
and from equations (\ref{15}) we have that $\gamma=4/3$ and
\begin{equation}    
\label{20.1}\Lambda_{2}\,=\,k\,-\,{\frac{2}{3}}\mathcal{B}\,\,,
\end{equation}
and the general behavior of the solutions (\ref{20}) do not change since we
have the condition $\mathcal{B}\not = 0$. However, we have to notice that the
square root demands attention in order to not obtain imaginary scale factors
for $k=0,\pm1$. But this is a trivial task to substitute in (\ref{20}) the
values of $\Lambda_{2}$ obtained in (\ref{20.1}) with different $k$. We will
not have any new physical feature in this specific mathematical analysis. The
equation of state is $p={\frac{1}{3}}\,\rho$, a radiation-dominated epoch as
said above.


Next we show a solution that does not have constraints in
$\Lambda_{1}$ and $\Lambda_{2}$ besides those given by the equations (16)%

\begin{equation}   
a(t)\,=-{\frac{\sqrt{\Lambda_{{2}}}\left(  \sinh\left(  \mathit{c_1}+\mathit{c_2}%
\,t\right)  \right)  ^{2}}{\mathit{c_2}}.}%
\end{equation}

If we set $\Lambda _{1}=-1$ and $\Lambda _{2}=1$ the solution has a singularity in $\tau $ $=-0.1279126267.$ Figure 5 show this result.






\begin{figure}[!h]
\begin{center}
\rotatebox{-90}{\includegraphics[width=0.35\textwidth]{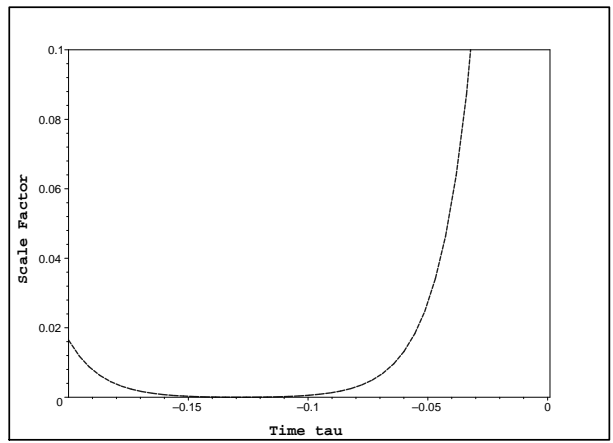}}
\caption{Scale factor versus $\tau ,$ equation (26), with $\Lambda _{1}=-1$ and $\Lambda _{2}=1$ $\left( k=-1\text{ and }\gamma =0\right)$ with singularity in $\tau $=$-0.1284233275$. (Units $H_{0}^{\text{ \ }-1}$ )}
\end{center}
\label{fig5}
\end{figure}

\bigskip
\section{Final considerations}

The equation (\ref{12}) was studied in \cite{oc,ln,ot,abdel,av2,cw,cops,mp} and it can also  
be obtained theoretically from some assumptions from quantum cosmology.  It is not in conflict with experimental observations and with the inflationary scenario.  From quantum cosmology arguments it is more convenient to use the scale factor instead of the age $t$ of the Universe \cite{ot}.

In this paper we have computed some analytical solutions for the scale factor since the effective cosmological constant $\Lambda$ is time varying according to an inverse square law in the scale factor.  This inverse square law can be justified based on the principles of quantum cosmology.  Since such time variation of $\Lambda$ leads to creation of matter and/or radiation, we believe that the formulation of a scale factor analysis is of current interest both theoretical and experimental.  Another motivation can be found in the strict relation between the cosmological constant problem and the problem of the age of the Universe treated in the light of quantum cosmology.

There is a great literature about different functions for the cosmological constant.  Specifically concerning a function like in the equation (\ref{12}) the existing published papers do not bring any analytical expressions for the scale factor itself as a building term in $\Lambda$.

As we are interested in a general analytical behavior of $\Lambda$, numerical solutions of (\ref{14}) are out of the scope of this work.  Therefore we found three analytical solutions.   Other alternatives varying the parameter $\Lambda_1$ in (\ref{14}) brought singularities which demands numerical efforts.

Our main motivation is to show simply that there is another possible and different approach concerning the cosmological constant in order to obtain some progress in some questions about the Universe.  Although the model proposed by Chen and Wu in equation (\ref{12}) are compatible with the astrophysical data, it alters the predictions of the standard model for the matter-dominated epoch.  

As a future perspective it can be investigated, with the solutions obtained here, some aspects of inflation like the current value of the deceleration parameter and the parameter $\Omega_0$, namely, to obtain its value and test if they are compatible with the standard model demands.  Note that the knowledge of the form of the scale factor, as well known, can describe if an Universe is an expanding one or not through the line element.

An investigation about the age of the Universe in different scenarios as well as the equations of state, with the results obtained in this paper, is a work in progress.

\section{Acknowledgments}

The authors would like to thank Prof. Felipe T. Falciano for helpfull discussions; 
Funda\c{c}\~ao de Amparo \`a Pesquisa do Estado do Rio de Janeiro (FAPERJ) and Conselho Nacional de Desenvolvimento Cient\' ifico e Tecnol\'ogico (CNPq) (Brazilian agencies) for financial support.

\pagebreak
\section*{Appendix}
\setcounter{equation}{0}

\subsection*{Full List of Scale Factor Solutions.}

\bigskip

\hspace{0.5cm} Quadratic solution:

\begin{equation}
a(t)\,=\,-\,2\,{\frac{\Lambda_{2}}{c_{1}}}\,+\,{\frac{c_{1}}{4}%
}\,(t\,+\,c_{2})^{2}\,\,,
\end{equation}

Exponential type solutions:

\begin{align}
a_{1}(t)\,=\,{\frac{1}{4c_{1}}} \left[  e^{c_{1}(t+c_{2}%
)}\,-4\,\Lambda_{2}\,e^{-c_{1}(t+c_{2})}\right]  \,\,,\nonumber\\
a_{2}(t)\,=\,{\frac{1}{4c_{1}}} \left[  e^{-c_{1}(t+c_{2})}\,-4\,\Lambda
_{2}\,e^{c_{1}(t+c_{2})}\right]  \,\,,
\end{align}

If $\Lambda_{1}=1,$ with square root type solution:%

\begin{equation}
a(t)\,=\,\pm\sqrt{{\frac{c_{1}}{\Lambda_{2}}}\,-\,\Lambda
_{2}(t\,+\,c_{2})^{2}}\,\,,
\end{equation}

Solutions with arbitrary $\Lambda_{1}$ and $\Lambda_{2}$:

\bigskip%

\begin{equation}    
a\left(  t\right)  =c_{3}\pm\frac{\sqrt{-\Lambda_{{1}}\Lambda_{{2}}}\left(
c_{1}+c_{2}\,t\right)  }{\Lambda_{{1}}c_{2}}.
\end{equation}

If $\Lambda_{1}=\frac{1}{n}$ ($n\in N$), where $n$ is a positive even number, the
solution is%

\begin{equation}   
a\left(  t\right)  =\sqrt{-\Lambda_{{1}}\Lambda_{{2}}\left(  t+c_{2}\right)
^{2}}%
\end{equation}

\bigskip If $\Lambda_{1}=-1,$ we have the following solutions%

\begin{equation}
a\left(  t\right)  =\pm{\frac{\sqrt{-\Lambda_{2}}\cosh\left(  C_{2}%
+C_{3}\,t\right)  }{C_{3}}},
\end{equation}%

\begin{equation}   
a\left(  t\right)  =\mp\frac{1}{2}\,{\frac{\sqrt{-\Lambda_{2}}}{C_{3}}\pm
\frac{\sqrt{-\Lambda_{2}}\left(  \cosh\left(  C_{2}+C_{3}\,t\right)  \right)
^{2}}{C_{3}}},
\end{equation}%

\begin{equation}   
a\left(  t\right)  =\mp{\frac{\sqrt{-\Lambda_{2}}\cosh\left(  C_{2}%
+C_{3}\,t\right)  }{C_{3}}\pm}\frac{4}{3}\,{\frac{\sqrt{-\Lambda_{2}}\left(
\cosh\left(  C_{2}+C_{3}\,t\right)  \right)  ^{3}}{C_{3}}},
\end{equation}%

\begin{equation}   
a\left(  t\right)  =\pm{\frac{\sqrt{\Lambda_{2}}\sin\left(  C_{2}%
+C_{3}\,t\right)  }{C_{3}}},
\end{equation}%

\begin{equation}   
a\left(  t\right)  =\mp\frac{1}{2}\,{\frac{\sqrt{\Lambda_{2}}}{C_{3}}\pm
\frac{\sqrt{\Lambda_{2}}\left(  \sin\left(  C_{2}+C_{3}\,t\right)  \right)
^{2}}{C_{3}}},
\end{equation}%

\begin{equation}   
a\left(  t\right)  =\mp{\frac{\sqrt{\Lambda_{2}}\sin\left(  \mathit{\_C2}%
+\mathit{\_C3}\,t\right)  }{\mathit{\_C3}}\pm}\frac{4}{3}\,\,{\frac
{\sqrt{\Lambda_{2}}\left(  \sin\left(  \mathit{\_C2}+\mathit{\_C3}\,t\right)
\right)  ^{3}}{\mathit{\_C3}}},
\end{equation}%

\begin{equation}    
a\left(  t\right)  =\pm{\frac{\sqrt{\Lambda_{2}}\cos\left(  \mathit{\_C1}%
+\mathit{\_C2}\,t\right)  }{\mathit{\_C2}}},
\end{equation}%

\begin{equation}    
a\left(  t\right)  =\mp\frac{1}{2}{\frac{\sqrt{\Lambda_{2}}}{\mathit{\_C2}}%
\pm\frac{\sqrt{\Lambda_{2}}\left(  \cos\left(  \mathit{\_C1}+\mathit{\_C2}%
\,t\right)  \right)  ^{2}}{\mathit{\_C2}}},
\end{equation}%

\begin{equation}    
a\left(  t\right)  =\mp{\frac{\sqrt{-\Lambda_{2}}\cos\left(  \mathit{\_C1}%
+\mathit{\_C2}\,t\right)  }{\mathit{\_C2}}\pm\frac{4}{3}\,\frac{\sqrt
{-\Lambda_{2}}\left(  \cos\left(  \mathit{\_C1}+\mathit{\_C2}\,t\right)
\right)  ^{3}}{\mathit{\_C2}}},
\end{equation}%

\begin{equation}    
a\left(  t\right)  =\pm{\frac{\sqrt{-\Lambda_{2}}\cosh\left(  \mathit{\_C1}%
+\mathit{\_C2}\,t\right)  }{\mathit{\_C2}}},
\end{equation}%

\begin{equation}    
a\left(  t\right)  =\mp\frac{1}{2}\,{\frac{\sqrt{-\Lambda_{2}}}{\mathit{\_C2}%
}\pm\frac{\sqrt{-\Lambda_{2}}\left(  \cosh\left(  \mathit{\_C1}+\mathit{\_C2}%
\,t\right)  \right)  ^{2}}{\mathit{\_C2}}},
\end{equation}%

\begin{equation}    
a\left(  t\right)  =\mp{\frac{\sqrt{-\Lambda_{2}}\cosh\left(  \mathit{\_C1}%
+\mathit{\_C2}\,t\right)  }{\mathit{\_C2}}\pm}\frac{4}{3}\,\,{\frac
{\sqrt{-\Lambda_{2}}\left(  \cosh\left(  \mathit{\_C1}+\mathit{\_C2}%
\,t\right)  \right)  ^{3}}{\mathit{\_C2}}},
\end{equation}%

\begin{equation}    
a\left(  t\right)  =\pm{\frac{\sqrt{\Lambda_{2}}\sin\left(  \mathit{\_C1}%
+\mathit{\_C2}\,t\right)  }{\mathit{\_C2}}},
\end{equation}%

\begin{equation}    
a\left(  t\right)  =\mp\frac{1}{2}\,{\frac{\sqrt{\Lambda_{2}}}{{\mathit{\_C2}%
}}\pm\frac{\sqrt{\Lambda_{2}}\left(  \sin\left(  \mathit{\_C1}+\mathit{\_C2}%
\,t\right)  \right)  ^{2}}{\mathit{\_C2}}},
\end{equation}%

\begin{equation}    
a\left(  t\right)  =\mp{\frac{\sqrt{\Lambda_{2}}\sin\left(  \mathit{\_C1}%
+\mathit{\_C2}\,t\right)  }{\mathit{\_C2}}\pm}\frac{4}{3}\,\,{\frac
{\sqrt{\Lambda_{2}}\left(  \sin\left(  \mathit{\_C1}+\mathit{\_C2}\,t\right)
\right)  ^{3}}{\mathit{\_C2}}},
\end{equation}%

\begin{equation}    
a\left(  t\right)  =\pm{\frac{\sqrt{\Lambda_{2}}\sinh\left(  \mathit{\_C1}%
+\mathit{\_C2}\,t\right)  }{\mathit{\_C2}}},
\end{equation}%

\begin{equation}    
a\left(  t\right)  =\mp\frac{1}{2}\,{\frac{\sqrt{-\Lambda_{2}}}{\mathit{\_C2}%
}\mp\frac{\sqrt{-\Lambda_{2}}\left(  \sinh\left(  \mathit{\_C1}+\mathit{\_C2}%
\,t\right)  \right)  ^{2}}{\mathit{\_C2}}},
\end{equation}%

\begin{equation}    
a\left(  t\right)  =\mp{\frac{\sqrt{\Lambda_{2}}\sinh\left(  \mathit{\_C1}%
+\mathit{\_C2}\,t\right)  }{\mathit{\_C2}}}\mp\frac{4}{3}\,\,{\frac
{\sqrt{\Lambda_{2}}\left(  \sinh\left(  \mathit{\_C1}+\mathit{\_C2}\,t\right)
\right)  ^{3}}{\mathit{\_C2}}}.
\end{equation}

\bigskip If $\Lambda_{1}=-2,$ we found out Jacobi elliptic type solutions

\bigskip%

\begin{equation}    
a\left(  t\right)  =\pm\frac{1}{2}\,{\frac{\sqrt{2\,\Lambda_{{2}}%
}{\mathit{JacobiDN}}\left(  \mathit{\_C2}+\mathit{\_C3}\,t,\sqrt{2}\right)
}{{\mathit{\_C3}}}},
\end{equation}%

\begin{equation}    
a\left(  t\right)  =\pm{\frac{\sqrt{-\Lambda_{{2}}}\mathit{JacobiNC}\left(
\mathit{\_C2}+\mathit{\_C3}\,t,\frac{1}{2}\sqrt{2}\right)  }{\mathit{\_C3}}},
\end{equation}%

\begin{equation}    
a\left(  t\right)  =\pm\frac{1}{2}\,{\frac{\sqrt{-2\Lambda_{{2}}%
}\mathit{JacobiND}\left(  \mathit{\_C2}+\mathit{\_C3}\,t,\sqrt{2}\right)
}{\mathit{\_C3}}},
\end{equation}%

\begin{equation}    
a\left(  t\right)  =\pm\frac{1}{2}\,{\frac{\sqrt{-2\Lambda_{{2}}%
}\mathit{JacobiNS}\left(  \mathit{\_C2}+\mathit{\_C3}\,t,i\right)
}{\mathit{\_C3}}},
\end{equation}%

\begin{equation}    
a\left(  t\right)  =\pm\frac{1}{2}\,{\frac{\sqrt{2}\sqrt{\,\Lambda_{{2}}%
}\mathit{JacobiSN}\left(  \mathit{\_C2}+\mathit{\_C3}\,t,i\right)
}{\mathit{\_C3}}}.
\end{equation}

\end{document}